\documentclass[aps,pre,twocolumn,float]{revtex4}

\usepackage{amsmath,bm,epsfig}
\def \ed {\end{document}}
\def\Fbox#1{\vskip1ex\hbox to 8.5cm{\hfil\fboxsep0.3cm\fbox{%
  \parbox{8.0cm}{#1}}\hfil}\vskip1ex\noindent}  

\newcommand{\eq}[1]{(\ref{#1})}
\newcommand{\Eq}[1]{Eq.~(\ref{#1})}
\newcommand{\Eqs}[1]{Eqs.~(\ref{#1})}
\newcommand{\Fig}[1]{Fig.~\ref{#1}}
\newcommand{\Ref}[1]{Ref.~\cite{#1}}

\def\be{\begin{equation}}\def\ee{\end{equation}}
\def\bea{\begin{eqnarray}}\def\eea{\end{eqnarray}}
\def\bse{\begin{subequations}}\def\ese{\end{subequations}}
\newcommand{\BE}[1]{\begin{equation}\label{#1}}
\newcommand{\BEA}[1]{\begin{eqnarray}\label{#1}}
\newcommand{\BSE}[1]{\begin{subequations}\label{#1}}

\let \nn  \nonumber  \newcommand{\br}{\\ \nn}
\newcommand{\BR}[1]{\\ \label{#1}}

\let \= \equiv \let\*\cdot \let\~\widetilde \let\^\widehat \let\-\overline
\let\p\partial

  \def\1{\bm1} 

\def\<{\left\langle}    \def\>{\right\rangle}
\def\({\left(}          \def\){\right)}
 \def \[ {\left [} \def \] {\right ]}

\renewcommand{\a}{\alpha}\renewcommand{\b}{\beta}\newcommand{\g}{\gamma}
\renewcommand{\d}{\delta}
\newcommand{\e}{\epsilon}\newcommand{\ve}{\varepsilon}
\renewcommand{\o}{\omega} 
\renewcommand{\L}{\Lambda}

\def\r{\rho}\def\k{\kappa}

\newcommand{\B}[1]{{\bm{#1}}}
\newcommand{\C}[1]{{\mathcal{#1}}}    

\renewcommand{\sb}[1]{_{\text {#1}}}  
\newcommand{\Sp}[1]{^{^{\text {#1}}}} 
\def\Sb#1{_{\scriptscriptstyle\rm{#1}}}

\begin{document}

\title{ Bottleneck crossover between classical and quantum superfluid turbulence  }
\author{Victor S. L'vov$^*$,      Sergei V. Nazarenko$^\dag$
and Oleksii Rudenko$^*$}
  \affiliation{$^*$Department of Chemical
Physics, The Weizmann Institute of Science, Rehovot 76100, Israel}

\affiliation{ $^\dag$University of Warwick, Mathematics Institute,
Coventry, CV4 7AL, UK}

\begin{abstract}
We consider superfluid turbulence near absolute zero of temperature
generated by
classical  means, e.g. towed grid
or rotation but not by counterflow. We argue that such turbulence
 consists of a {\em polarized} tangle of mutually
interacting vortex filaments with quantized vorticity.
For this system
we predict and describe a bottleneck accumulation of the energy
spectrum at the classical-quantum crossover scale $\ell$. Demanding
the same energy flux through scales, the value of the energy at the
crossover  scale should exceed  the Kolmogorov-41 spectrum by a large
factor    $\ln^{10/3} (\ell/a_0)$   ($\ell$   is   the mean
intervortex distance and $a_0$ is the vortex core radius) for the
classical   and quantum spectra to be matched in value.
One of the important consequences of the bottleneck is that
it causes the mean vortex line density to be considerably higher
that based on K41 alone, and this should be taken into account in
(re)interpretation of  new (and old) experiments as well as in further theoretical studies.

\end{abstract}

\maketitle

\section*{Introduction}

Turbulence in superfluid liquids, such as $^4$He and $^3$He at very
low temperatures, is an intriguing physical problem by itself
because it comprises a system where the classical physics gets
gradually transformed into the quantum one during the energy cascade
from large to small scales~\cite{1,2}. Recently  renewed broad
interest to this subject has been motivated by an impressive
progress in experimental techniques and new
  results, which has led  in (at least) conceptual
understanding of classical and quantum limits of the superfluid
turbulence,  see,
e.g.~\cite{exp0,exp,exp1,exp2,exp3,exp4,araki,stalp,VinenNiemela,
Vin03-PRL,svistunov,cn,Naz_kelvin}. Our paper in turn, attempts to
shed light on physics of the superfluid turbulence behavior in the
intermediate region near the classical-quantum crossover scale.
We will see that transition of the turbulent energy cascade from the
classical to the quantum scales is accompanied by transition from
strong hydrodynamic to weak wave turbulence with a bottleneck
stagnation at the crossover scale.

Generally, the superfluid turbulence near zero temperature (for an
introduction, see~\cite{VinenNiemela,1,2}) can be viewed as a tangle
of quantized vortex lines. If turbulence is produced by classical
means and not by a counterflow then at the scales much greater than
the mean intervortex distance $\ell$,   the vortex discreteness is
unimportant, so that the superfluid turbulence has essentially a
classical character described by the Kolmogorov-41 (K41)
approach~\cite{VinenNiemela}. As we will see below, vortex lines in
K41 state are polarized, i.e. tend to  be co-directed and organized
in bundles. Since there is no viscosity or friction in a superfluid
liquid near zero temperature, the classical energy cascade proceeds
down the spectrum to the scale of order $\ell$ without dissipation,
where the vortex discreteness and quantization effects become
important. Even though some negligible part of the energy is lost,
for example, by
 radiation of phonons generated due to slow
vortex motions and intermittent vortex reconnections, the dominant
part of the energy proceeds to cascade below the scale $\ell$ by
means of nonlinearly interacting Kelvin waves
\cite{VinenNiemela,Vin03-PRL,5,svistunov}, which were theoretically
predicted in the 19th century~\cite{19th} and first experimentally
observed by Hall~\cite{Hall}. We emphasize that the fact that the
turbulence is produced by classical means  is important here,
because resulting polarization inhibits further vortex reconnections
and prevents rapid fragmentation into vortex loops with sizes
smaller
 than $\ell$. Thus, the main cascade carrier below scale $\ell$ will
be Kelvin waves which are generated by both slow vortex filament
motions and fast (but more rare and localized) vortex reconnection
events. Such reconnections produce sharp bends on the vortex lines
and, therefore, generate a broader range of wavelengths than the
slow vortex motions. However, the spectrum of the reconnection
forcing decays with the wavenumber $k$ sufficiently fast, and could
effectively be thought as a large-scale source of Kelvin waves
located at the crossover~\cite{Naz_kelvin}. Traditionally, the K41
spectrum is assumed to maintain its shape all the way down to the
crossover scale, which, due to such an assumption, is calculated
based on the K41 spectrum \cite{VinenNiemela}.

\begin{figure} \vskip -.3cm
\includegraphics[width=8 cm]{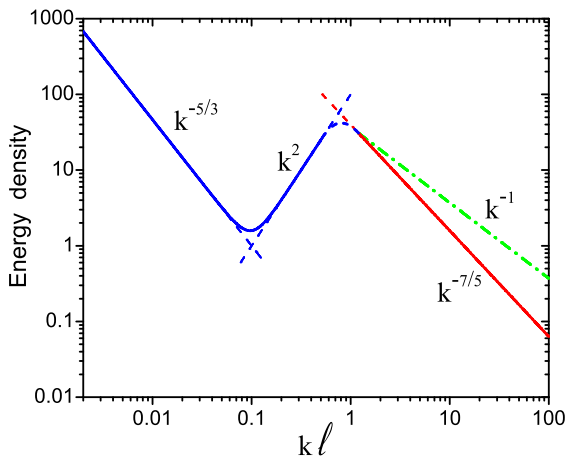}
  \caption{
(Color online). The energy spectra $\C E_k$ in the classical, $k < 1/\ell$, and quantum,
$k > 1/\ell$, ranges of scales. Two straight blue (dark gray) lines in the
classical range indicate the pure K41 scaling $\C E_k\Sp
{K41}\propto k^{-5/3}$, \Eq{19},  and the pure thermodynamic scaling
$\C E_k \propto k^{2}$. For the quantum range, the red (gray) solid line
indicates the Kelvin wave cascade spectrum, \Eq{16}   (slope
$-7/5$), whereas the green (light gray) dash-dotted line marks the spectrum
corresponding to the non-cascading part of the vortex tangle energy
(slope $-1$). }\label{f:1}
\end{figure}
 In   this paper, we demonstrate that, in contrast to the
traditional viewpoint,  the classical turbulent  spectrum cannot be
matched to its quantum counterpart  at the same value of the energy
flux because this flux requires much stronger levels of turbulence
to be able to propagate through scales in sparse distributions of
quantized vorticity. This
  leads to a bottleneck accumulation of the energy spectrum near
the crossover scale which, in turn,   significantly changes the
position of the crossover, $\ell$, (see \Fig{f:1}) and the
relationship between the energy flux and vorticity, which have
widely been used in interpretation of  experimental results. Notice,
that the phenomenon of bottleneck accumulation between two
energy-flux spectra of different nature is not peculiar to the
superfluid turbulence and may occur, for example, in the atmosphere,
ocean and magnetosphere.

  \section{ Polarization of the vortex tangle}

It is important that turbulence  we consider in this paper is
generated by classical means, e.g. by a towed grid \cite{stalp} or
by rotation~\cite{exp,exp1,exp2}, but not by a counterflow. In the
later case the vortex tangle would be unpolarized and neither we
would expect K41 spectrum for the scales greater than $\ell$  (K41
is polarized, see below) nor would we expect Kelvin waves to be
important for the scales below $\ell$ (reconnections would be more
important, see \cite{06Nem}). On the other hand, polarization of the
vortex tangle allows to shape large-scale vortex motions
characteristic to the K41 cascade, and it also inhibits local
reconnections and makes Kelvin waves a dominant vehicle for the
turbulent energy cascade toward the small scales. Thus, let us
consider the phenomenon of vortex polarization in greater detail.

Intuitively, polarized vortex tangle can be viewed as a set of vortex
bundles, so that in each  bundle the  vortex filaments have the same
preferential direction. The simplest way to achieve such a polarization
is to subject the system to an external rotation or shear.
However, as we will see below, even isotropic and homogeneous turbulence
can be, and often is,  polarized.

 Let us formalize this picture by giving a
mathematical definition of the vortex tangle polarization. Consider
a circular disk  or radius $R$ with randomly selected position of
its center and its orientation in the 3D space. The velocity
circulation over the contour of this disc, $\Gamma (R)$, is
obviously equal to the quantum circulation $\kappa$ (see Eq.~(\ref{H})) multiplied by
the difference between the number of vortices crossing the disk in
the positive and the negative direction with respect to the normal
to the disk,
\be \label{1}\Gamma (R) = \kappa (N_+ - N_-). \ee
 The totally unpolarized
system is represented by a vortex tangle in which every vortex line
consists of a chain of small uncorrelated segments (as in
\cite{06Nem}). In this case
 the disc crossings would be completely random, and the
mean value of $\Gamma^2$ would be determined from the central-limit
theorem. Namely, if the sign of each crossing is completely
random and statistically independent of all the other crossings
then the total circulation $\Gamma$ has zero mean and
the standard deviation equal to the standard deviation for the
circulation of
an individual crossing $\kappa^2$ times the total number of
terms in the sum (i.e. the number of crossings),
\BSE{2}\BE{2a}
\langle \Gamma^2 \rangle = \kappa^2 \langle N_+
+ N_- \rangle \sim \kappa^2 (R/\ell)^2, \ee
 where $\ell$ is the
mean intervortex distance. We will say that this state has zero
polarization, $P=0$. Thus, the polarization $P$ can be defined as a
degree of deviation from this unpolarized state. For example,
 in the completely polarized system all vortex lines
would be in perfectly aligned state, e.g. $N_- =0$ and $N_+>0$, so
that
 \BE{2b}
 \langle \Gamma^2 \rangle = \kappa^2 \langle
N_+^2\rangle \sim \kappa^2 (R/\ell)^4\,,  \ee\ese 
We will say that in
this state $P=1$. Let us now define polarization $P$ by
interpolating between these two limits. Namely, we will assume
that the system is in a scaling state such that
\BSE{3} \BE{3a}\langle \Gamma^2 \rangle = \kappa^2 \langle
N_+^2\rangle \sim \kappa^2 (R/\ell)^\sigma\,  \ee
with  some constant index $\sigma$. Then for this state
the polarization is defined as
\BE{3b} P=\sigma/2 -1\ . \ee\ese
 Note that in principle one can
have a vortex system in which $P<0$, e.g. an ordered grid
structure composed of alternating positive and negative vortices.
However, the alternating periodic structures are unstable and would
quickly break up due to reconnections.

Polarization of turbulent states with power law spectra is considered
in Appendix B describing three different cases. For very
steep spectra $P=1$, for very shallow spectra (including the thermodynamic
state) $P=0$, and there is a window of intermediate spectra (including K41)
for which $P$ depends on the spectral slope and, therefore, contains
a nontrivial information about the turbulent scalings.
For K41 turbulence we have
 \BE{4} 
 \langle \Gamma^2 \rangle_{K41} \sim
\ve^{2/3} R^{8/3}\ . \ee
In this case $\langle \Gamma^2 \rangle$ can also be obtained from the
dimensional analysis.
Thus, for K41 turbulence we have
$\sigma =8/3$ and polarization $P=1/3$.

Therefore, the vortex tangle associated with the K41 cascade state
is polarized. Note that  in presence of bottleneck (described below)
there will also be a contribution of the thermalized part of the
spectrum. However, this part is much less
that the K41 contribution for large $R/\ell$. On the other hand, at
scale $R \sim \ell$ (and obviously for $R < \ell$) the notion of
polarization becomes vague and useless, so one should not attempt to
find $P$ for these scales.

\begin{figure} \vskip -.3cm
\includegraphics[width=8 cm]{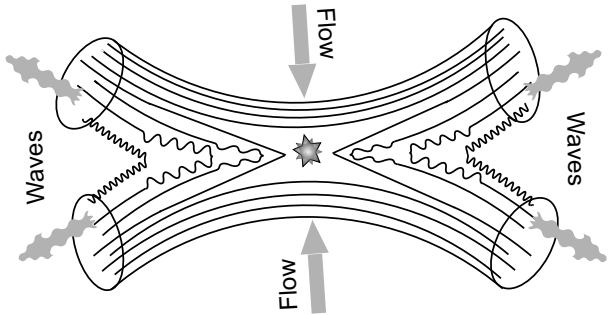}
  \caption{A sketch of typical reconnection of vortex lines in polarized
vortex tangles. A clash of two vortex bundles results in a localized
reconnection region. Kelvin waves generated by the reconnections
propagate away from the localized reconnection region and spread in
space along the vortex lines.} \label{f:2}
\end{figure}

Significant polarization associated with K41 cascade at large scales
leads to  grouping of the adjacent vortex lines into bundles with
predominantly parallel orientation which obviously inhibits
reconnections and which selects Kelvin waves to be the dominant
carrier of the downscale energy cascade. This picture is
self-consistent because, as we will see later, only weak Kelvin
waves are needed to carry the energy cascade of the same strength as
in the large-scale K41 part. Associated with such weak waves small
bending angles will not allow the adjacent (collinear) vortex
lines to approach each other and reconnect. On the other hand, it
should be emphasized that the dominance of Kelvin waves over the vortex
reconnections still remains a hypothesis, even though a very
plausible one. In this picture, the regions of reconnections are
intermittent and limited to locations where two vortex bundles
clash, see Fig.~\ref{f:2}. Kelvin waves generated by such localized
reconnections will spread in space along the vortex lines into the
vortex bundles. Therefore the resulting wave distributions will be
much less localized in space than the reconnections. Note also that
the reconnections in this picture do not lead to a creation of vortex
loops of size $\ell$ or less and, therefore, cannot trigger a
cascade of further fragmentation of such loops, as it would be the
case in unpolarized tangles in counterflow experiments \cite{06Nem}.
Assuming that the large-scale dynamics of  strongly
 polarized systems is similar to the classical flows described
by Euler equations,
one could imagine a classical prototype
process in which reconnections will intermittently occur
in locations of (yet to be proven to exist) singularities of the Euler equations.
One can also see a clear analogy with reconnections of magnetic field
lines in MHD with Alfven or whistler waves being similar to Kelvin waves.

  \section{ Kinetics of interacting Kelvin waves}

Let us describe the statistics of Kelvin waves on thin vortex filaments,
and their role as a carrier of the  energy cascade at the scales
less than the interline separation $\ell$.
 Here we
briefly  overview  the results of Kozik and Svistunov (2004) on this
problem~\cite{svistunov} (hereafter referred to as KS-04) with
modifications and clarifications, particularly keeping an explicit
account for the logarithmic factors which will be important
for the effects found in our work.
 The motion
of the tangle of quantized vortex lines can be described by the
Biot-Savart equation (BSE) \cite{1,2} for the evolving in time
radius vector of the vortex line element $\B s(\xi,t)$, depending on
the arc lengths $ \xi $ and time $t$. When the typical interline
spacing $\ell$  is large in the sense $ \L\=\ln(\ell/a_0)\gg 1$
($a_0$ is the vortex core radius), this equation can be simplified by
the so-called local induction approximation (LIA) \cite{3}. Both BSE
and its LIA can be written in the Hamiltonian form~\cite{5}:
$$\displaystyle
i\kappa
\dot{w} = \delta H\{w,w^*\} / \delta w^*,$$
where $ w(z,t)\= x(z,t)+
i\,y(z,t)$ with $x$ and $y$ being small   distortions of the
almost straight vortex line along the Cartesian $z$-axis. The BSE
and LIA Hamiltonians are: 
\BSE{H}\BEA{BSE}
 H \Sp{BSE}\!\! &=& \frac{\kappa^2}{4\pi} \int \! \! {\{ 1+{\rm Re} \,
   [w'^*(z_1)w'(z_2)]\} \, dz_1 dz_2 \over
     \sqrt{(z_1-z_2)^2+|w(z_1)-w(z_2)|^2}}\,,~~~~~\BR{LIA}
 H \Sp{LIA}\!\! &=& \frac{\kappa^2 \L }{2\pi} \int \! \! \sqrt {\{ 1+|
    w'(z)|^2 }  \, dz\,, \\ 
\nonumber \k &\=& {2\pi \hbar}/m\,,
\eea\ese
where primes denote the $z$-derivatives, $\k$  is the quantum of
velocity circulation, and $ m$ is the particle mass. Without the
cut-off, the integral in $H\Sp{BSE}\!$, \Eq{BSE}, would be
logarithmically divergent with the dominant contribution given by
the leading order expansion of the integrand in small $z_1-z_2$,
that corresponds to $H\Sp{LIA}\!$, \Eq{LIA}.

It is well known that LIA represents a completely integrable system
and it can be reduced to one-dimensional nonlinear Schrodinger (NLS)
equation by Hasimo transformation \cite{hasimoto}.
 However, it is the complete integrability of LIA that makes it insufficient for
describing the energy cascade and which makes it necessary to
consider the next order corrections within the BSE model.

Assuming that the  Kelvin wave amplitudes are small with respect to
their wavelengths, i.e. $ w'\ll 1 $ (the self-consistency of this assumption is
checked by an estimate of the nonlinearity parameter, see \Eq{xi-k}), we can expand the
Hamiltonians~\eq{H} in powers of ${w'}^2$: $H=H_0+H_2+H_4+H_6+ \ldots
$ Next step is to consider a periodic system with the period length
$\C L$ ($\C L
 \to \infty$ to be taken later) and to use  the
Fourier representation $  w(z,t)= \kappa^{-1/2} \sum_k a(k,t) \exp
(ikz)\, $ in terms of which the Hamiltonian equation  takes  the
canonical form
\BE{can} \nonumber
 i\,   {\p a(k,t)}/{\p t } = {\p \C H\{a,a^*\} }/{\p a(k,t)^*}
\ee
 with a new Hamiltonian $ \C  H\{a ,a ^*\}=
 H\{w,w^* \} / \C L = \C H_2 + \C H_4 + \C H_6 + \dots$ With $a_j\=a(k_j,t)$:
 \BEA{eqA}\nn  && \hskip -0.6 cm  \C H_2 =\sum_k \o_k\, a(k)
a^*\!(k) \,, \  \C H_4 =\frac14\sum_{12,34}T_{ 12,34}\,a_1a_2a_3^*
a_4^*\,, \BR{Exp}
 && \hskip -0.6 cm \C H_6
=\frac1{36}\sum_{123,456}W_{123,456}\,a_1a_2a_3 a_4^*a_5^* a_6^*\,.
 \eea
Here $\o_k$ is the Kelvin wave frequency and  interaction amplitudes
$T_{12,34}$, $W_{123,456}$ are functions of $k_1\dots k_4$ and
$k_1\dots k_6$, correspondingly. Summations over $k_1\dots k_4$ in
$\C H_4$ and over $k_1\dots k_6$ in $\C H_6$ are constrained by
$k_1+k_2=k_3+k_4$ and $k_1+k_2+k_3=k_4+k_5+k_6$, correspondingly.
One gets for functions in the Hamiltonians: 
\BSE{LIA-int}
\BEA{7a} 
\omega\Sp{BSE}_k&=& {\kappa \L (k)}\, k^2/ {4\pi}\,, \quad\ \L (k)\=
\ln (1/k a_0)\,,\br T_{1,2;3,4}\Sp{BSE}&=&   k_1k_2k_3k_4  [\L (k
\sb {ef} ) + F_{1,2;3,4}]/4\pi\,,\br 
W_{1,2,3;4,5,6}\Sp{BSE}&=& 9 k_1k_2k_3k_4k_5k_6 [\L (k \sb {ef} ) +
F_{1,2,3;4,5,6}]/ 32 \pi \kappa\,, \br
\br 
 \omega\Sp{LIA}_k&=& {\kappa \L\,}k^2/{4\pi} \,, \quad  T_{1,2;3,4}\Sp{LIA}=  {\kappa \L}\, k_1
k_2 k_3  k_4/{4\pi}, \BR{7b}
 W_{1,2,3;4,5,6}\Sp{LIA}&=&
 9\, \L  k_1k_2k_3k_4k_5k_6/{32 \pi \kappa}\ . 
\eea\ese  
Here $k\sb{ef}$  is a mean value of wave vectors in the game and all
functions $F$ are of the order of unity, they depend of
the ratios of involved $k_j$ to $k\sb{ef}$.

 It is well known that four-wave dynamics in one-dimensional case with
dispersion laws $\o_k\Sp{BSE}$ or $\o_k\Sp{LIA}$ is absent because
the conservation laws of energy and momentum allow only trivial
processes with $k_1=k_3$,   $k_2=k_4$, or $k_1=k_4$,   $k_2=k_3$.
However, nontrivial six-wave scattering processes of $3\to3$ type
are allowed. For weakly interacting waves this dynamics can be
described in terms of correlation functions $ \displaystyle \<
|a(k,t)|^2  \> = \C L^{-1} \,  n(k,t)$, with the help of a classical
six-wave kinetic equation \cite{svistunov}, shown below for  the
continuous limit $\C L \to \infty$ and $n_j\=n(k_j,t)$:
\bea\nn
\frac{\p n_k}{\p t}\!\!&=&\!\! \frac{\pi}{12}\!\!
\int \!\!|\~W_{k,1,2;3,4,5}|^2 \left [ \C N _{3,4,5;k,1,2 }- \C N
_{k,1,2;3,4,5}\right] \BR{KE}
&& \times \d(\o_k \!+\!\o_1 \!+\!\o_2 \!-\!\o_3 \!-\!\o_4 \!-\!\o_5) \, \br 
&& \times \delta(k \!+\! k_1 \!+\!k_2 \!-\!k_3 \!-\!k_4 \!-\!k_5)\,
dk_1 dk_2\!\dots dk_5\,,\br
 && \hskip -1.35cm \C N_{1,2,3;4,5,6} \=
  n_1n_2 n_3(n_4n_5+n_4n_6+n_5n_6)\ .
\eea
Here $\~W$ is the \emph{full} interaction amplitude, that includes
the bare 6-wave amplitude $W$ and 72 contributions of the 2nd order
in the 4-wave amplitudes  of the order of $T_{k1,23}^2/\o_k$.
Notably, that LIA   has infinitely many integrals of motion due to
its complete integrability. These integrals totally preserve system
from dynamical evolution: $\p n(k,t)/\p t\=0$ for any $n(k)$
distribution. With 6-wave kinetic \Eq{KE}, this is possible only if
$\~W_{k,1,2;3,4,5}\Sp{LIA}=0$ on the resonant manifold
$k+k_1+k_2=k_3+k_4+k_5$, $ \o_k+\o_1+\o_2=\o_3+\o_4+\o_5$. This
means that the leading contribution to $\~W_{k,1,2;3,4,5}\Sp{BSE}$,
proportional to $\L$ (that coincides with
$\~W_{k,1,2;3,4,5}\Sp{LIA}$) also vanishes due to cancelations of
the leading contribution to $W_{k,1,2;3,4,5}\Sp{BSE}$ with that
originating from the perturbative terms. Remaining terms in
$\~W_{k,1,2;3,4,5}\Sp{BSE}$ can be presented   as follows
\BE{12}
\~W_{k,1,2;3,4,5}\Sp{BSE}=    k_1k_2k_3k_4k_5k_6 \, \Phi
_{1,2,3;4,5,6}\big / 4\pi \k\,,\ee
where some dimensionless function $\Phi _{1,2,3;4,5,6}$ is   of the
order of unity and depends only on mutual ratios of  $k$-vectors
$k_1\dots k_6$. This estimate differs from \Eq{7a} for
$W_{k,1,2;3,4,5}\Sp{BSE}$ by absence of the large factor $\L$.

The kinetic equation~\eq{KE} written for a single vortex filament
has a stationary   solution~\cite{svistunov} with a constant energy
density (per unit length) flux $\e$.  We reformulate this ``KS-04"
spectrum to the 3D vortex tangle system in terms of the rate of
energy density (per unit mass) in the 3D space, $\ve =
\e/\rho\,\ell^2$  ($\rho$ is the fluid density):
\begin{equation}
n_k \simeq {(\ell^2 \ve)^{1/5}} \k^{2/5}\, |k|^{-17/5}\,, \ \ \
\mbox{KS-04 spectrum.}
\label{17/5}
\end{equation}

It should be mentioned, though, that the theory of the cascade energy spectrum (KS-04)
was derived with an assumption that vortex lines in the tangle are not rectilinear
and non-interacting. In the present work, having in mind that reconnections are dominated
 by the mean intervortex distance, we silently assumed that the interactions and
non-rectilinearity of vortex lines become unimportant at small scales.

\section{ Warm cascades   in    hydrodynamic turbulence}

 The
energy density per unit mass for Kelvin waves of small amplitude is
\BE{e-dens}\nonumber
\C E= L \int \C \o_k n_k {d\,k}/{2\pi}\= \int \C E_k dk/2\pi,
\ee
where   $L\simeq \ell^{-2}$ is the vortex line density per unit
volume and  $\C E_k$ is one-dimensional energy density in the
$k$-space.  Together with \Eq{17/5} this gives
\BE{16} 
\C E_k\simeq  \L
  \big(  \k^7 \ve \big  / \,  \ell^8 \big)^{1\!/5}\, |k|^{-7/5} .
\ee
Note that the parameters $\ve$ and $\ell$ in \Eq{16} are mutually dependent. Their relation follows from the
expression for the mean vorticity in the system of quantum
filaments, $\<|\o|\>\simeq \k L \simeq \k \ell^{-2}$, where $\<|\o|\>$ is dominated by
the classical-quantum crossover scale and its estimate is usually
based on the K41
 spectrum,
\BE{19}
\C E_k\Sp {K41}\simeq \ve^{2/3} |k|^{-5/3}\,, \ee 
which gives
\BE{est-K41}\nonumber
\<|\o|\>^2 \sim \int^{1/\ell} k^2 \C E_k\Sp {K41} dk \sim \ve^{2/3}
\ell^{-4/3}, \ \mbox{or}\ \ \ve \sim \k^3 / \ell^4.~
\ee
However, this estimate is rather unprecise because the K41 spectrum
cannot be matched to the Kelvin wave spectrum
\Eq{16} at the crossover scale and, as it is explained below,
there exists a bottleneck. But since the bottleneck is on the
classical side of the spectral range, and
the mean vorticity is still dominated by the crossover scale,
one can find the correct relation between $\ve$ and $\ell$ based
on \Eq{16} instead of K41. This gives:
\BE{est-our}\nonumber
\<|\o|\>^2 \!\sim {1/\ell}^3 \C E_k|_{k=1\!/\ell} \sim
 \L\, (\k^7\ve / \ell^{16} )^{1\!/5},
\ \ \mbox{or}\ \ \ve \sim {\k^3 \big /  \L^5 \ell^4}.
\ee
This estimate is
different from the standard one based on K41 by a large factor of $\L^5$.

Now, from \Eqs{16} and (\ref{19}), one can find the ratio of quantum
and classical (K41)
 spectra of turbulence at the crossover scale
$k\simeq 1/\ell$: 
\BE{21} 
\C E_{1\!/\ell}\Big / \C E_{1\!/\ell}\Sp {K41}\simeq \L^{10/3}\gg 1\ .
\ee
This ratio shows that quantum turbulence of Kelvin waves requires
much higher level of energy (by factor $\L^{10/3}$) in order to
provide the same rate of the energy flux (and the same rate of
the energy dissipation) than in the  hydrodynamic turbulence of
classical fluid. The main reason of that is the ``rigidity" of the
vortex filaments which is reflected by factor $\L_k$ in \Eq{7a} in
the Kelvin wave frequency. This contributes a factor of $\L^{8/3}$
into the ratio~\eq{21}. The remaining factor of $\L^{2/3}$
originates from the fact that any one-dimensional system of
interacting Kelvin waves described by the Bio-Savart equation is
close to the fully integrable LIA system, in which the dynamics of
wave amplitudes is absent. Thus, in order to have the same value of
the energy flux and continuity of the spectrum at the crossover
scale, there must be a bottleneck pile-up of the classical spectrum
near this scale by the factor of $\L^{10/3}\!$, and this will be
described by a ``warm cascade'' solution in what follows.
\vskip .1cm

As we explained, the energy flux carried by classical hydrodynamic
turbulence with K41 spectrum~\eq{19}  cannot fully propagate through
the crossover region. Therefore, hydrodynamic motions with larger
scales (smaller wave-vectors) will increase their energy up to the
level $\C E_{1/\ell}$, required for Kelvin waves to maintain the
same energy flux. As the result, for $k\le 1/\ell$ the spectrum of
hydrodynamic turbulence, $\C E\Sp{HD}_k\!$, will not have the K41
scale-invariant form $\C E\Sp{K41}_k\!$ given by \Eq{19}. To get a
qualitative understanding of resulting bottleneck we will use so
called ``warm cascade'' solutions found in \cite{cn}. These
solutions follow from the Leith-67 differential model for the energy
flux of hydrodynamical turbulence,
\BE{22} \ve_k=- {1  \over  8 } \,
   \sqrt{|k|^{13}F_k}\  {d F_k  \over   d k}  \,,
\quad F_k\= \frac{\C E\Sp{HD}_k\!\!}{k^2}\,,  \ee 
where  $F_k$ is the 3-dimensional spectrum of turbulence. Generic
spectrum with a constant energy flux can be found as the solution to
the equation ~$\ve_k=\ve$: 
\BE{23}
 F_k =  \Big[ \frac{24\ve }{11 |k|^{11/2}}+
\Big(\frac{T}{\pi \rho}\Big)^{3/2}\Big]^{2/3}\ .
 \ee 
The large $k$ range describes a thermalized  part of the spectrum
with equipartition of energy characterized by an effective
temperature $T$, namely, $T/2$ of energy per a degree of freedom,
thus, $F_k = T \big/ \pi \r$ and $ \C E_k = T k^2 \big/ \pi \r$. At
low $k$, \Eq{23} coincides with K41 spectrum, \Eq{19}.

 This ``warm cascade'' solution describes reflection of K41
cascade and stagnation of the spectrum near the bottleneck scale
which, in our case, corresponds to the classical-quantum
crossover scale. To obtain the spectrum in the classical range of
scales, it remains now to find $T$ by matching \Eq{23} with the
value of the Kelvin wave spectrum at the crossover scale $\C E_k
\sim \k^2/\ell$. This gives $T/\r \sim \k^2  \ell \sim (\k^{11}
/\L^5 \ve)^{1/4}.$

Obviously, the transition between the classical and the quantum
regimes is not sharp and in reality we should expect that a gradual
increase of the role of the self-induced wave-like motions of individual
vortex lines
with respect to the collective classical-eddy type of motions
of the vortex bundles. Thus, the high-wavenumber
 part of the thermalized range is likely to be wave rather than eddy dominated.
However, the energy spectrum for this part would still be of the same
$k^2$ form corresponding to the thermal energy equipartition.
This picture relies on the assumption (justified below) that the self-induced wave
motions have small amplitudes and, therefore, do not lead to
reconnections.

The resulting spectrum including both the classical, the quantum and
the crossover parts,  is shown on Fig.~1 as a log-log plot.
Important, that at $k > 1/\ell$ in addition to the cascading energy
associated with  Kelvin waves, there is also energy associated with
the tangle of vortex filaments (shown on Fig.~1 by a green
dash-dotted line). The energy spectrum of this part $\sim |k|^{-1}$,
which is simply a spectrum associated with a singular distribution
of vorticity along 1D curves in the 3D space \cite{araki} and does
not support a down-scale cascade of energy. The cascading and
non-cascading parts have similar energies at the crossover scale,
that is the wave period and the amplitude are of the order of the
characteristic time and size of evolving background filaments. In
other words, the scales of the waves and of the vortex ``carcass''
are not separated enough to treat them as independent components.
This justifies the matching of the classical spectrum at the
crossover scale with the Kelvin wave part alone ignoring the
``carcass'' which is valid up to an order-one factor. This also
justifies the way of connecting the ``carcass'' spacing $\ell$ to
the cascade rate $\ve$.

\section{Weakness of turbulence at and below scale $\ell$}

In principle, turbulent fragmentation cascade into decreasingly
 small vortex loops can be an alternative to Kelvin waves
as mechanism of the energy transfer below scale $\ell$.
Dimensionally, one can obtain spectrum $\propto |k|^{-1}$ which
corresponds to such a cascade~\cite{5,16} (and which occidentally
coincides with the non-cascading ``carcass'' spectrum discussed
above). The probability of such small-scale reconnections depends on
statistics of vortex orientations and can be estimated only in the
simplest case of totally unpolarized vortex tangle by adopting a
model in which every vortex line consists of short-correlated
pieces, see e.g.~\cite{06Nem}. This model is relevant for turbulence
produced by a thermal counterflow but not to the case of polarized
turbulence produced by classical means. Definitely, in the case
when, at the microscopic level, the vortex lines are preferably
parallel (presumably this is the case for superfluid turbulence in
the rotating tank~\cite{exp,exp1,exp2}), the reconnection
scenario~\cite{5,16} is irrelevant, as we assumed in our approach.

This picture is supported by estimation of the nonlinearity parameter
$\xi_k$ through comparison of the nonlinear frequency shift
$\Delta\omega_k$ with the frequency itself: \BE{xi-k}
\xi_k \=
 \frac{\Delta \omega_k}{\omega_k} \simeq
 \frac{T_{k,k;k,k}\Sp{BSE}\, n_{\!k}\, k}{\omega\Sp{BSE}_k} \simeq
 \frac{1}{\Lambda\, (k\ell)^{2/5}}\, , \ee
which is obtained using \Eqs{7a}, (\ref{17/5}) and our estimate $\ve
\sim {\k^3 \big /  \L^5 \ell^4}$.
 Now we see that at the cross-over
scale the nonlinearity is small: 
\begin{equation}\nonumber
\xi_{1\!/\ell} \simeq 1/\Lambda\simeq 1/\ln{\!(\ell/a_0)} \ll 1\ . 
\end{equation}
Correspondingly, characteristic values of the bending  angle $\alpha$ associated with
Kelvin waves are also small,
\begin{equation}\label{al}
\alpha \sim \sqrt{\xi} \sim
1/\sqrt{\Lambda} \ll 1\ .
\end{equation}
 Hence, the KS-04 weakly nonlinear spectrum
should dominate the Svistunov-95 reconnection spectrum~\cite{5}.
Indeed, the mean wave amplitude at scale $\ell$ is $\sim \alpha
\ell$, i.e. too small for the adjacent vortex lines to ``touch" and
reconnect. This proves self-consistency of the picture of the cascade
carried by interacting Kelvin waves without reconnections, but we
emphasize that this picture assumes polarization of the vortex
system on which Kelvin waves propagate.

Similarly, in the thermalized region of scales the mean bending
angle can be estimated as $\alpha \sim \sqrt{\xi} \sim
(k \ell)^3 /\sqrt{\Lambda} \ll 1.$ Thus,
the self-induced vortex line motions gradually arising
from the eddy-like collective motions in the thermalized part
take form of weakly nonlinear Kelvin waves.
The nonlinearity of Kelvin waves grows with $k$ in the thermalized
part reaching its peak at the crossover scale and decreasing
in the Kelvin cascade range.

\section*{ Summary and Discussion}

In this paper, we suggested the following
route for the development of the energy
cascade:
 K41 $\rightarrow$ "Warm Up" $\rightarrow$
KS-04 spectrum with a very pronounced Bottleneck effect. This scenario
is relevant for {\em polarized} vortex systems resulting from
forcings of a classical type, e.g. a towed grid or rotation, but not
relevant to unpolarized vortex tangles produced by thermal
counterflows. In our arguments we relied on the fact that
polarization suppressed the reconnection-fragmentation cascade.
Classically produced K41 turbulence is indeed polarized. However,
its polarization is not perfect and at this time we cannot exclude
that in some specific cases,
 the reconnection dynamics can suppress the
bottleneck accumulation of energy.

In this paper we predicted that the
bottleneck on the classical-quantum crossover scale  amplifies
the spectrum at this scale by a large  factor of $\L^{10/3}$ with
respect to K41. Correspondingly, the corrected estimate for the
crossover scale which takes this bottleneck into account  is $\ell
\sim (\k^3 / \L^{5} \ve)^{1/4}$, which is
 $\L^{5/4}$ times smaller  than the standard estimate
based on K41. Yet another way to reformulate the same thing
would be to say that the effective viscosity $\nu^{\,\prime}$ is
reduced by a large factor of $\Lambda^5$, i.e.
\begin{equation}
\nu^{\,\prime} \approx \kappa /\Lambda^5
\label{nu}
\end{equation}
 (see e.g. \cite{VinenNiemela}
for definition of $\nu'$ and explanation of its meaning).

A comment is due about locality of the transition between the
classical turbulence and the Kelvin wave cascade. Due to a sharp
kink-like nature of the vortex reconnections generating Kelvin waves
(see Fig.~\ref{f:2}), one might think that the energy is injected
into
 the Kelvin wave cascade over a wide range of wavenumbers
 (associated with a Fourier
analysis of the kink), and conclude that
  the energy spectrum in the quantum
region should differ from  $k^{-7/5}$.
  However, as it was shown in \Ref{Nazar-Arxiv},
 the Fourier transform of the kink decays with $k$ fast enough
 for the direct cascade scaling to  dominate. In other words,
  the reconnection forcing appears to be more or less equivalent to a
low-frequency forcing at the intervortex scale $\ell$.

To describe the shape of the bottleneck spectrum, we used the ``warm
cascade'' solution previously obtained in \cite{cn} based on the
Leith-67 differential model \Eq{22}, as it is the simplest model
which can provide a clear qualitative understanding of the
bottleneck phenomenon.
Clearly, the differential model \Eq{22} exaggerates locality of the
interactions of scales in real Navier-Stokes turbulence where the
main contribution to evolution of $\C E\Sp{HD}_k$ originates from a
wider range of comparable scales $q\sim k$. Some authors claim that
extended interaction triads with $q$ between $k/A$ and $Ak$ (with
$A\sim 10$) are most important~\cite{Orz}. If so, the transient
region between K41 and the thermodynamically equilibrium spectra can
be wider than the one predicted by the differential approximation
\Eq{22}. To account for this effect, one can use a more
sophisticated turbulence closure based on an integral rather than
the differential equation, e.g. one of the traditional closures such
as the Direct Interaction Approximation (DIA) or ``Eddy-Damped
Quasi-Normal Markovian" (EDQNM), as it was done in ~\Ref{Bos}, or
even simpler closure, suggested in   Appendix A.

In this paper, we did not consider the effects of the mutual friction
between the normal and superfluid components, thereby restricting
our consideration to low temperatures (e.g. below 1K for $^4$He).
At higher temperatures the dissipation of energy by the
mutual friction can exceed the energy transfer to Kelvin waves,
which would make our analysis and conclusions inapplicable.
This seems to be the case, for example in experiments described
in \cite{stalp}.

At lower
temperatures there is a clear lack of experiments on turbulence
generated by classical means. In this respect, one could mention
the $^3$He experiment on turbulence
generated by a vibrating wire at Lancaster \cite{exp0}
the authors of which found the value
of the effective viscosity $\nu' = 0.2 \kappa$ which appears
to be much greater than our prediction (\ref{nu}).
On the other hand, to obtain this value the authors used an estimate
for the integral (energy containing) scale to be equal to the
thickness of the turbulent region, $d=1.5$ mm which in our opinion may not be the
case for the oscillating grid setup.
Lacking direct measurements of $d$, we could get
guidance from the oscillating grid experiments in
classical fluids, see e.g. \cite{eidelman} where the following
estimate for the integral scale is given,
$$
d= 0.25 (S/M)^{1\!/2} z,$$
where $M$ is the mesh size, $S$ is the amplitude of
oscillations, $z$ is the distance from the grid.
Taking the Lancaster parameters, $M= 50\,\mu$m, $S=1\,\mu$m, $z=1.5$ mm,
 we get
$d = 50\,\mu$m. Estimating $\nu' $ with this value of the integral
scale would give $\nu' \sim 2 \times 10^{-4}\kappa$, which would be consistent
with the small $\Lambda^{-5}$ coefficient in (\ref{nu}).
However, it is not possible to be more conclusive one way or another
without more direct measurements of the turbulent parameters in this case.

 A lucky exception appears to be a new $^3$He
experiment on rotation generated turbulence, in which the
bottleneck phenomenon appears to be important in understanding
the observed propagation speed of the turbulent-laminar interface,
see \cite{nature} for detailed explanations.

In the final stages of modifying our paper according to the referee comments,
our attention was drawn to a new preprint \cite{KS-Arxiv} where an alternative
picture of the crossover turbulence was presented with bottleneck being
prevented by reconnections. The authors argued that the self-induced part
of the vortex line velocity becomes larger than the classical collectively
produced velocity in the vortex bundle already at the scale $r_0 \sim \Lambda^{1/2} l_0
\gg l_0$. From this they concluded that the polarized vortex bundles move
randomly with respect to each other which leads to their random reconnections.
In this respect we would like to re-emphasize our view that the fast
self-induced motions take form of rapidly-oscillating Kelvin waves rather than
of a chaotic motion of vortex bundles. Moreover, as expressed in our estimate (\ref{al}),
these Kelvin waves must have very small bending angles ($1/\Lambda^{1/2} $ or less)
in order for the six-wave Kelvin cascade to carry the same flux as in the K41 (large-scale)
part of the spectrum.
Such small bending angles are insufficient for the neighboring lines within a particular bundle
to approach each other and to reconnect neither in the thermalized
nor in the cascade range of scales. Thus, reconnections are limited to rather
small volumes in-between of colliding large-scale bundles.

On the other hand, as we have already said in this paper, polarization of K41 turbulence
is not perfect and we may expect reconnections to play some role which would lead to
certain modifications of the bottleneck phenomenon described in this paper.
Relative role of the reconnections versus the Kelvin wave cascades
 is also likely to depend on the particular way of turbulence excitation.
In particular, we may expect further reinforcement of the polarization, and therefore
stronger suppression of reconnections, in rotating systems and in systems with
a strongly sheared mean flow.

\section*{Acknowledgements}
The authors are thankful to S.~Nemirovskii, B.~Svistunov and W.F.~Vinen for
fruitful discussions. V.L. is grateful to G.~Volovik, M.~Krusius and
V.~Eltsov for involving him into discussion of ongoing
experiment~\cite{nature} on turbulence in superfluid He$^ 3$, that shed
light on the bottleneck problem. The work is partially supported by
the US-Israel Binational Science Foundation and by Finberg
fellowship for O.R. In part, this work was done during the EPSRC
sponsored
 Warwick Turbulence Symposium 2005-2006,
which comprised a series of workshops on different aspects of turbulence
and a year-long visitor programme.

\appendix

\section{Simple turbulent closure}

Here we propose a new model which could be viewed as the simplest
(minimal) integral closure to be used in the future for an improved
description of the transitional bottleneck region. The model
comprises in writing the collision term St$\{F_k\}$ in  energy
spectrum balance $ {\p F_k}/{\p t}=\mbox{St}\{F_k\}$ as follows:
 \BEA{24}
  \mbox{St}\{F_k\}&\simeq&
   \int  _{ -\infty}^\infty
  \frac{q^2 d   q \, p^2 d   p
 \, \d (  k +  q+   p)}{2\pi\,k^2\, (k^2+q^2+p^2)}\br
 &&\hskip -1cm \times {  k \big [   k\, F_q F_p +
    q\, F_k F_p +    p\, F_q F_k    \big ]}\big / (\g_k+\g_q+\g_p)\,,
\eea 
where $  k$,   $  q$, and $  p$ are one-dimensional vectors varying
in the interval $(-\infty, + \infty)$,  and $\displaystyle
  \g_k \=  \sqrt {|k|^5 F_k  } $ represents eddy-turnover frequency.
The model~\eq{24}   differs from   EDQNM   by replacement of $d^3 q
\, d^3 p \, \d^3( \B k +\B  q+ \B  p) $ with 3-dimensional vectors
$\B  k$,   $\B  q$, and $\B p$~~ by~ $ q^2 d q \, p^2 d   p
 \  \d (  k +  q+   p)/ (k^2+q^2+p^2)$ with
  one-dimensional vectors, by a simpler form of $\g _k$,
and by one-dimensional version of
   the interaction amplitude
    ($V^{\a\b\g}_{  k   q   p} \Rightarrow    k$).

The model~\eq{24}  satisfies all general closure requirements: it
conserves energy, $\int k^2 \mbox{St}\{F_k\} dk =0$  for any  $F_k$,
and St$\{F_k\}=0 $ on the thermodynamic equilibrium spectrum
$F_k=$const and  on the cascade spectrum $ F_k\propto |k|^{-11/3}$.
Importantly, the integrand in \Eq{24}  has the correct  asymptotical
behavior at the limits of small and large $q/k$ as in the
sweeping-free Belinicher-L'vov representation, see \Ref{BL}. This
means that our model adequately  reflects contributions of the
extended interaction triads and thus should be useful in the future
for a quantitative description of the transient region between
turbulent spectra with the thermodynamic and the energy-flux
equilibria.

\section {Velocity circulation and Polarization in turbulence}

In this Appendix we will calculate the velocity circulation
$\Gamma$ over a circular contour of radius $R$ in classical
turbulence with a power-law spectrum.
Let the second order velocity correlation function
(3D spectrum) in the ${\bf k}$-space for isotropic
homogeneous turbulence be
\begin{equation}\label{F-k}
  F_k = C{\Sb F}\frac{{v{^2\Sb T}} {k}^{\,x-3}_{*}}{k^{\,x}}\,,
\end{equation}
where $C{\Sb F} = 2 \pi^2 |x-3|$,
$v{\Sb T}$ is the rms velocity in turbulence
and $k_{*}$ is a  wavenumber of truncation  from
above for $x < 3$ (e.g. for the thermodynamic equilibrium
with $x=0$) and from below for $x > 3$ (e.g. for K41\
 turbulence with $x={11/3}$).
Such a truncation is necessary for $v{\Sb T}$ to be
finite, and we will see below that the $x=3$
boundary also separates different types of the scaling
behavior of the velocity circulation.

Consider the circulation
\begin{equation}\label{Gamma-circ}
  \Gamma = \int_R \omega_n d^2 r\,,
\end{equation}
where ${\bf \omega} = {\bf \nabla} \times {\bf v}$ is vorticity, and the
integral is taken around an arbitrary circle of radius R. Then
\begin{equation}\label{Gamma2-circ}
  \<\Gamma^2\> = \int_R\int \<\omega_{1,n}\,\omega_{2,n}\> d^2 r_1 d^2 r_2\,,
\end{equation}
Due to isotropy of the turbulence we may approximate $\<\omega_{1,n}\omega_{2,n}\> = \frac13 \<{\B \omega}_{1}\cdot{\B \omega}_{2}\>$. With ${\bf r}_{12} \= {\bf r}_{1} - {\bf r}_{2}$ and
$\<{\B \omega}_{k_1}\cdot{\B \omega}_{k_2}\> = 2\,
k^2_1\, (2\pi)^3\delta({\bf k}_1+{\bf k}_2)\, F_{k_1}$, we have 
\begin{eqnarray}\label{ww-1}
  \<{\B \omega}_{1}\cdot{\B \omega}_{2}\> =
\int\!\!\!\int\!\! \frac{d^3 k_1 d^3 k_2}{(2\pi)^6}\
e^{i ({\bf k}\cdot{\bf r}_1 +{\bf k}\cdot{\bf r}_2)}
\<{\B \omega}_{k_1}\cdot{\B \omega}_{k_2}\> ~~~~~&& \\ 
  = 2\!\!\int\!\! k^2 F_k\, e^{i\, {\bf k}\cdot{\bf r}_{12}} \frac{d^3 k}{(2\pi)^3} 
  = \frac{C{\Sb F}}{\pi^2} \frac{v{^2\Sb T}\,k_*^{\,x-3}}{r_{12}}
\!\!\int^\infty_0{\!\frac{\sin(k\,r_{12})}{k^{\,x-3}}dk}\ .\nonumber &&
\end{eqnarray}
When $3 < x < 5$, this integral converges and we have
\begin{equation}\label{ww-3x4}
  \<{\B \omega}_{1}\cdot{\B \omega}_{2}\> = 2 |x-3|
\frac{v{^2\Sb T}}{r^2_{12}}\! \({k_*r_{12}}\)^{{x-3}}\!\left\{-{\cal G}(4-x)
\sin \frac{\pi x}{2} \right\},
\end{equation}
where ${\cal G}(x)$ is the Gamma-function. Substituting this expression into
(\ref{Gamma2-circ}) and integrating we have
\begin{equation}\label{ww-3x4}
  \<\Gamma^2\> = C_x
\frac{v^2_T}{k_*^2} \({k_*R}\)^{{x-1}},
\end{equation}
where $C_x$ is an order-one constant (whose analytical  dependence  on index $x$
is very complicated).

For $x<3$, the integral in (\ref{ww-1})
diverges at the upper limit and, therefore,
has to be truncated at the maximum wavenumber which, in this case,
is $k_*$. We have,
\begin{eqnarray}\label{lt3}
  \<{\B \omega}_{1}\cdot{\B \omega}_{2}\>
  = \frac{C{\Sb F}}{\pi^2} \frac{v{^2\Sb T}\,(k_* r_{12})^{x-3}}{r_{12}^2}
\!\!\int^{k_* r_{12}}_0{\!\frac{\sin y}{y^{x-3}}dy}\ . &&
\end{eqnarray}
The integral in this expression can be found in terms
of the special functions whose asymptotical behavior can be readily
obtained. This way one can show that
the  correlator $\<{\B \omega}_{1}\cdot{\B \omega}_{2}\>$
decays in $ r_{12}$ sufficiently fast, so that
for $k_* R \gg 1$ one can pass in the integral (\ref{Gamma2-circ})
to the symmetric variables ${\bf r}_+=\frac12({\bf r}_1+{\bf r}_2)$
and ${\bf r}_{12}={\bf r}_1 -{\bf r}_2$, use the polar coordinates
and replace the upper integration limit for $r_{12}$ with
infinity,
\begin{eqnarray}\label{Gamma2-lt3t}
  \<\Gamma^2\> &=& \frac{2 \pi^2 R^2}{3}  \int_0^\infty \<{\B \omega}_{1}\cdot{\B \omega}_{2}\> r_{12}d r_{12} \,.\nonumber\
\end{eqnarray}
Substituting $\<{\B \omega}_{1}\cdot{\B \omega}_{2}\>$ from
(\ref{lt3}) and integrating  we have
 \begin{equation}\label{lt3a}
  \<\Gamma^2\>
   = \frac {4 \pi^4} {3} {v{^2\Sb T}\,R^{2}}.
\end{equation}
Interestingly, this expression is independent of both
the spectrum exponent $x$ and of the cutoff $k_*$.
This fact has a simple physical interpretation.
Suppose that the correlation length of vorticity field
in turbulence $\ell_\omega$ is short so that
$\ell_\omega \ll R$. Then the circle or radius $R$ embraces
$N=R^2/\ell_\omega^2$ random and effectively independent
vortex tubes each having radius $\sim \ell_\omega$ and
circulation $\gamma_l \sim v{\Sb T} \ell_\omega$. Because of the statistical
independence of these tubes, $\<\Gamma^2\> $ can be found
using the Central Limit Theorem, similarly to the way we
did it in the main text for the set of random quantized
vortex lines. We have
$$
\<\Gamma^2\> \sim \gamma_l^2 {N} = v{^2\Sb T} R^2,
$$
which coincides, up to an order-one numerical factor, with
expression (\ref{lt3a}). Note that dependence on $\ell_\omega$
has dropped out, which corresponds to independence of
expression (\ref{lt3a}) of  $x$ and  $k_*$.

For $x>5$, the integral in (\ref{ww-1})
 diverges at the lower limit and, therefore,
has to be truncated at the minimum wavenumber
which, in this case,
is again $k_*$. We have,
 \begin{equation}\label{gt5}
  \<{\B \omega}_{1}\cdot{\B \omega}_{2}\>
  = \frac{C{\Sb F}}{\pi^2} \frac{v{^2\Sb T}\,k_*^{2}}  {(x-5)}.
\end{equation}
As we see, this correlator is independent of
the distance $r_{12}$ which simply means that the
correlation length in this case is of order of the
maximal lengthscale $1/k_*$.
The integration in (\ref{Gamma2-circ}) in this case
reduces to the multiplication by the square
of the circle area, i.e.
\begin{equation}\label{gt5a}
  \<\Gamma^2\> = \frac{1}{3}
\frac{C{\Sb F}}{\pi^2} \frac{v{^2\Sb T}\,k_*^{2}}  {(x-5)} (\pi R^2)^2
   = \frac {2 \pi^2 (x-3)} {3(x-5)} {v{^2\Sb T} k_*^2\,R^{4}}.
\end{equation}
The $R^4$ scaling here coincides with the one obtained in the
main text for a bundle of perfectly aligned (polarized)
vortex lines. This is not surprising since the vorticity
correlation
length in this case $\sim 1/k_*$ which is much greater
than the contour size $R$.
Another interesting effect to note is that
$ \<\Gamma^2\>$ diverges for $x \to 5$.

Let us now summarize all the cases. We have
\begin{eqnarray}\label{Gamma2-est-2}
 x < 3 \, \hbox{(e.g. thermodynamic)}:  & \<\Gamma^2\>
   =& \frac {4 \pi^4} {3} {v{^2\Sb T}\,R^{2}},  \nonumber\\
 3<x < 5 \;\;\; \hbox{(e.g. K41)}:  & \<\Gamma^2\>  =& C_x
\frac{v{^2\Sb T}}{k_*^2} \({k_*R}\)^{{x-1}},\nonumber
   \\
 x >5 \; \; \hbox{(e.g. smooth field)}:  & \<\Gamma^2\> =&
 \frac {2 \pi^2 (x-3)} {3(x-5)} {v{^2\Sb T} k_*^2\,R^{4}}.\nonumber
\end{eqnarray}
For polarization $P$ (see definition in the main text) we have
respectively
\begin{eqnarray}\label{p-summary}
P = \left\{
  \begin{array}{ll}
    0\,, & x < 3\,, \\
    (x-3)/2\,, & x \in (3,5)\,, \\
    1\,, & x > 5\ .
  \end{array}
\right.
\end{eqnarray}

So, the thermal equilibrium state ($x = 0$) is not polarized at all, $P = 0$, whilst Kolmogorov turbulence
 ($x = 11/3$) is partially polarized, $P = 1/3$.


\begin{thebibliography}{99}
\bibitem{1} R. J. Donnelly, Quantized Vortices in He II (Cambridge
University Press, Cambridge, 1991)
\bibitem{2} Quantized Vortex Dynamics and Superfluid Turbulence, ed. by
C. F. Barenghi   et al., Lecture Notes in Physics  \textbf{571}
(Springer-Verlag, Berlin, 2001)
\bibitem{VinenNiemela} W.~F. Vinen and J.~J. Niemela,   J. Low Temp.
Phys.   {\bf 128}, 167 (2002)
\bibitem{exp0} S. N. Fisher, A. J. Hale, A. M. Gu\'{e}nault, and G. R. Pickett,   Phys. Rev. Lett., \textbf{86}, 244 (2001)
\bibitem{exp} R. H\"anninen,  R. Blaauwgeers, V. B. Eltsov, A. P. Finne, M. Krusius, E. V. Thuneberg, and G. E. Volovik,   Phys. Rev. Lett. \textbf{90}, 225301 (2003)
\bibitem{exp1} A. P. Finne, T. Araki, R. Blaauwgeers, V. B. Eltsov, N. B. Kopnin, M. Krusius, L. Skrbek, M. Tsubota, G. E. Volovik,   Nature \textbf{424}, 1022  (2003)
\bibitem{exp2} D. I. Bradley, D. O. Clubb, S. N. Fisher, A. M. Gu\'{e}nault, R. P. Haley, C. J. Matthews, G. R. Pickett, V. Tsepelin, and K. Zaki, Phys. Rev. Lett. \textbf{95}, 035302 (2005)
\bibitem{exp3} V.B. Eltsov, A. P. Finne, R. H\"anninen, J. Kopu, M. Krusius, M. Tsubota, and E. V. Thuneberg, Phys. Rev. Lett. \textbf{96}, 215302 (2006)
\bibitem{exp4} D.I. Bradley, D. O. Clubb, S. N. Fisher, A. M. Gu\'enault, R. P. Haley, C. J. Matthews, G. R. Pickett, V. Tsepelin, and K. Zaki, Phys. Rev. Lett. \textbf{96}, 035301 (2006)
\bibitem{araki} T. Araki, M. Tsubota, and S. K. Nemirovskii, Phys. Rev. Letts.
\textbf{89}, 145301 (2002)


\bibitem{stalp} S.R. Stalp, J.J.Niemela, W.F.Vinen and R.J.Donnely,
 Physics of Fluids \textbf{14}, Issue 4, pp. 1377-1379 (2002)
\bibitem{svistunov} E.V. Kozik and B.V. Svistunov, Phys. Rev. Lett.
\textbf{92}, 035301 (2004)
\bibitem{Vin03-PRL} W. F. Vinen, M. Tsubota and A. Mitani, Phys. Rev. Lett.
 {\bf 91}, 135301, (2003)
\bibitem{cn} C. Connaughton and S. Nazarenko, Phys. Rev. Lett.
 {\bf 92}, 044501 (2004)
 \bibitem{Naz_kelvin}
S. Nazarenko, JETP Lets., \textbf{ 84}, 700 (2006)
 \bibitem{5} B. V. Svistunov, Phys. Rev. B \textbf{52}, 3647 (1995)
\bibitem{19th} W. Thomson, Philos. Mag. \textbf{10}, 155 (1880)
\bibitem{Hall} H. E. Hall, Proc. R. Soc. London, Ser. A \textbf{245}, 546 (1958)
\bibitem{06Nem} S.K. Nemirovskii, J. Low Temp. Phys. \textbf{142},
769 (2006).
\bibitem{3} K.W. Schwarz, Phys. Rev. B \textbf{31}, 5782 (1985) and
\textbf{38}, 2398 (1988)
\bibitem{hasimoto}  H. Hasimoto, J. Fluid Mech, \textbf{51} pp 477-485(1972)
\bibitem{16}  W.~F. Vinen, Phys. Rev. B \textbf{61}, 1410 (2000)
\bibitem{Nazar-Arxiv} S. Nazarenko, JETP Letters. \textbf{84},
 issue 11, 700-702 (2006).
\bibitem{Orz} S.A. Orzag, J. Fluid Mech. \textbf{41}, 363 (1970)
\bibitem{Bos} W. J. T. Bos and J.-P. Bertoglio, Phys. of Fluids, {\bf 18},
071701 (2006)
\bibitem{nature} V.B. Eltsov, A. Golov, R. de Graaf,
R. Hanninen, M. Krusius, V. L'vov, R.E. Solntsev, ``Propagation of
turbulent front in rotating superfluid", International workshop on
rotating superfluid turbulence, Propagation of turbulent front in
rotating superfluid turbulence, Jerusalem, April, 2007.
\bibitem{BL} V.I. Belinicher and V.S. L'vov,  Zh. Eksp. Teor. Fiz., \textbf{93}
 1269 (1987)
\bibitem{eidelman} A. Eidelman, T. Elperin, A. Kapusta, N. Kleeorin, A. Krein, and I. Rogachevskii,
Nonlinear Processes in Geophysics  \textbf{9} 201 - 205 (2002)
\bibitem{KS-Arxiv}  E. V. Kozik and B. V. Svistunov, (2007),\\ 	arXiv:cond-mat/0703047v2 [cond-mat.other]

\end{thebibliography}
\end{document}